# A Dual Model of Open Source License Growth

Gottfried Hofmann[1], Dirk Riehle[1], Carsten Kolassa[1], and Wolfgang Mauerer[2]

1  Friedrich-Alexander-Universität Erlangen-Nürnberg. Computer Science Department, Martenstrasse 3, 91058 Erlangen, Germany dirk@riehle.org,
WWW home page: http://osr.cs.fau.de/

2  Siemens AG, Corporate Research and Technologies,
San-Carlos-Str. 7. 91058 Erlangen, Germany
wolfgang.mauerer@siemens.com

**Abstract**. Every open source project needs to decide on an open source license. This decision is of high economic relevance: Just which license is the best one to help the project grow and attract a community? The most common question is: Should the project choose a restrictive (reciprocal) license or a more permissive one? As an important step towards answering this question, this paper analyses actual license choice and correlated project growth from ten years of open source projects. It provides closed analytical models and finds that around 2001 a reversal in license choice occurred from restrictive towards permissive licenses.

## 1  Introduction

Research on open source software (OSS) and development processes has gained significant momentum over the last decade. Landmark work was published by Lerner and Tirole in 2003 [1]. A meta-study was conducted by Aksulu and Wade in 2010 [2] to give an overview of the state of the research in the field. Yet many basic questions remain to be answered. One of them is the question of licensing.

When a project has the ability to chose its license freely, license choice is frequently controversial. The same applies to the situation where a project decides to switch from one license to another. Besides philosophical reasons to favor one type of license over another there is the concern whether the chosen license has an impact on the project's success.

Our research question is to understand the relationship between OSS licenses and project growth. In this paper we answer the question of which type of license do people prefer.

Roughly from the early 1960s to the early 1980s sharing of source code for computer programs was commonplace and conducted in an informal manner. This kind of collaboration happened in an academic setting. When commercial companies started to enforce intellectual property rights, the first open source licenses emerged as an effort to retain the collaborative environment by providing a legal framework.

Among the first of these initiatives was the Free Software Foundation (FSF) [3], which published a first version of the GNU General Public License in 1989 [4]. The GPL includes a clause that forces developers who make changes to the code to release their changes under the same conditions as the GPL. This property of the GPL led to the attribution of the GPL as a 'viral' [5] or 'reciprocal' license. Another term for this kind of licensing is 'copyleft'. For the remainder of this paper, licenses of this kind will be called 'restrictive'.

In 1988, two licenses were first published whose conditions were later coined 'copyfree' or 'permissive', namely the MIT license [6] and the BSD license [7][1]. Both do not require derived work to be licensed under the same terms[2], thus redistributing code for proprietary products is possible.

Later, licenses were created like the GNU Lesser General Public License (LGPL) that are less restrictive than the GPL-like licenses yet still not completely permissive. Projects that use those licenses are not subject of this analysis for the sake of simplicity.

Please note that both license types emerged roughly at the same time, so none of the two types used for the analysis here had a "head-start" over the others, see Fig. 1.

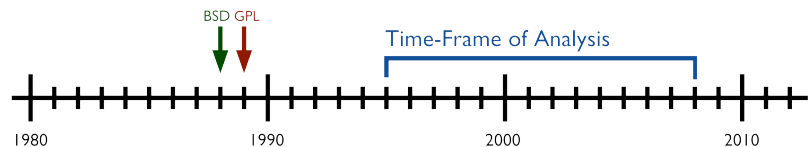

**Fig. 1.** Time-frame of the analysis.

This paper makes the following contributions:
- Two analytically closed models of the total open source growth binned by license-type are proposed.
- A validation of the models using statistical measures.
- An estimation of changing-points that separates the growth into two periods.

The rest of the paper is structured as follows. Section 2 reviews related work. Section 3 presents the data source and research method. Section 4 provides the discovered models and statistical validation. We discuss potential limitations in section 5 and present our conclusions in section 6.

---

[1]  Both licenses are available in multiple versions now, like the 2-clause, 3-clause and 4-clause BSD license or the X11 license.
[2]  Yet there are still restrictions like in the 'New BSD License' which does not permit advertising of derived products with the name of the licensor.

## 2 Related Work

Various studies have been conducted in the past to find out about the rationale behind a license choice. Sen, Subramian and Nelson [8] suggest that "OSS managers who want to attract a limited number of highly skilled programmers to their open source project should choose a restrictive OSS license. Similarly, managers of software projects for social programs could attract more developers by choosing a restrictive OSS license". Lerner and Tirole [9] argue that "Projects with unrestricted licenses attract more contributors". In contrast, Colazo and Fang [10] analyzed 44 restrictively- and 18 permissively-licensed projects from the SourceForge database. The restrictively-licensed projects had a significantly higher developer membership and coding activity.

In a series of articles [11] [12] [13], Aslett describes a recent trend in open source licensing that shows that the ratio of permissively- vs. restrictively-licensed projects is slowly shifting in favor of permissive licensing. Source for the data is both the Ohloh.net [14] database and FLOSSmole [15]. The time-frame of that analysis is from 2008 to 2011. We are not verifying these findings as the author looks at trends from 2008 onwards. This paper looks at the developments from 1995 to the middle of 2007 filling the gap left by Aslett.

Deshpande and Riehle [16] use 5122 active and popular open source projects from the Ohloh database as a sample and find that open source in both added SLoC per month and new projects per month shows in total exponential growth.

## 3 Data Source and Research Method

The sample source of this paper is a snapshot of the Ohloh.net [14] database dated March 2008. The Ohloh database has been collecting data of open source projects since 2005 from publicly visible revision control repositories. Since those repositories provide a history, the available data dates back as early as 1983 [17]. Yet data before 1995 was omitted as it was too sparse to be useful. Data after June 2007 was also omitted as it was not fully collected yet. According to Koch [18], revision control systems (RCS) are a very good source to study open source projects.

Our analysis is data driven: we are discovering existing characteristics in our data rather than starting off with a hypothesis and attempting to invalidate or validate it. We analyze how the total growth of open source projects can be correlated to the chosen type of license and provide closed-form models. We provide details of our final findings and list the models we tried to fit.

**3.1 Metrics Employed**

To measure growth of the size of projects, we use the metric Source Lines of Code (SLoC) added per month. A SLoC is a line in a commit (code contribution) that is neither empty nor a comment. According Herraiz et. al. [19], SLoC is a good metric to measure project growth. To show this they compared SLoC to various other common metrics of size (number of functions etc.) and complexity (McCabe's cyclomatic complexity, Halstead's length, volume, level etc.) of software projects and found a high correlation between them.

SLoC are calculated using the Unix diff command between two consecutive versions and then removing blanks and comments.

### 3.2 Growth

To determine the total growth of one license-type, all SLoC of all projects in a license-bin are added up in month-windows after removing the initial month. Removing the initial month is done to reflect the fact that the size at 'birth' of a project is not of interest when measuring growth. Thus the problems of forks and projects that started privately are also addressed[3].

We chose added SLoC per month because it represents all developers as opposed to choosing the number of projects started per month which would only represent those who started a project. Thus our approach is representative of the behavior of the entire developer population.

### 3.3 Distinction of License-Types

The model for permissive and restrictive licenses in this paper is based on the model proposed by Lerner and Tirole [9]. It was expanded by additional licenses that occur in the data set. All licenses are required to be approved by the OSI. Our sample contains 1861 projects in the category 'permissive' and 3257 projects in the category 'restrictive'. Projects offering both restrictive and permissive licenses are counted in both sets. Projects under 'mildly restrictive' or 'weak copyleft' licenses like the LGPL have been omitted for the sake of simplicity. Table 1 lists the number of occurrences in the sample.

The total number of projects included in the analysis is 5118 which is too large to list the individual projects in here. At its time, it constituted about 30% of all active open source projects.

---

[3]  Note that this does not account for the case when a project becomes open source but the history of the revision control system is preserved or when a fork imports the history, too.

**Table 1.** Licenses by Type. Multiple versions of a license are counted as one. For example GPL v1, v2 and v3 are listed as GPL only. Some projects have multiple licenses.

| Permissive | | Restrictive | |
|---|---|---|---|
| **License Name** | **Observations** | **License Name** | **Observations** |
| BSD | 730 | GPL | 3248 |
| MIT | 378 | CC-BY-SA | 24 |
| Apache | 479 | | |
| zlib/libpng | 26 | | |
| Public Domain[4] | 34 | | |
| Artistic License | 210 | | |
| Python license | 17 | | |
| Zope | 8 | | |
| Vovida | 1 | | |

## 4. Research Results

Fig. 2 shows the total added SLoC per month for the permissive and restrictive set. For the remainder of this paper, the data for the permissive set in each figure is on the left side and the restrictive set on the right. The blue curve is a smooth nonparametric fit obtained with the Loess method [20]. The curve shape is not influenced by a-priori considerations, it is solely data driven, and can be used as a visual aid in the comparison of descriptive models introduced below. The gray shaded area around the Loess curve represents the 95% confidence interval.

---

[4] Public Domain is considered a permissive 'license' in this paper.

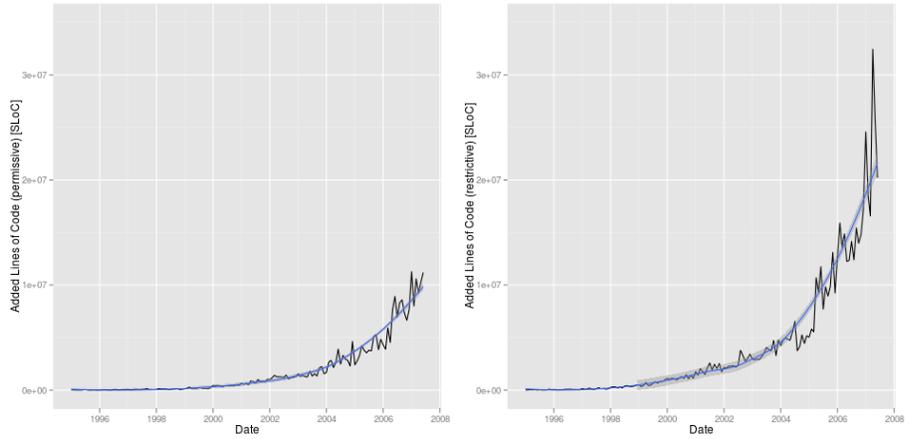

**Fig. 2.** Total SLoC added per month with blue Loess curve

The form of the monotonically growing Loess curve suggests the following model functions:

- Logistic (normal and 4-parameter)
- Gompertz
- Polynomials: Quadratic, Cubic
- Exponential

From the functions that returned a fit, we used Pearson's $r^2$ and visual inspection of the graphs to determine the best fit. For both sets the exponential model returned the highest Pearson's $r^2$ (0.960 for the permissive and 0.937 for the restrictive set) and best visual compliance. Equation (1) shows the formula for the exponential model.

$$y \sim y0 * \exp(a * x) \qquad (1)$$

As a remedy for the heteroscedasticity that can be seen in Fig. 2 we log-transformed the response. The graphs with Loess curve in blue are shown in Fig. 3.

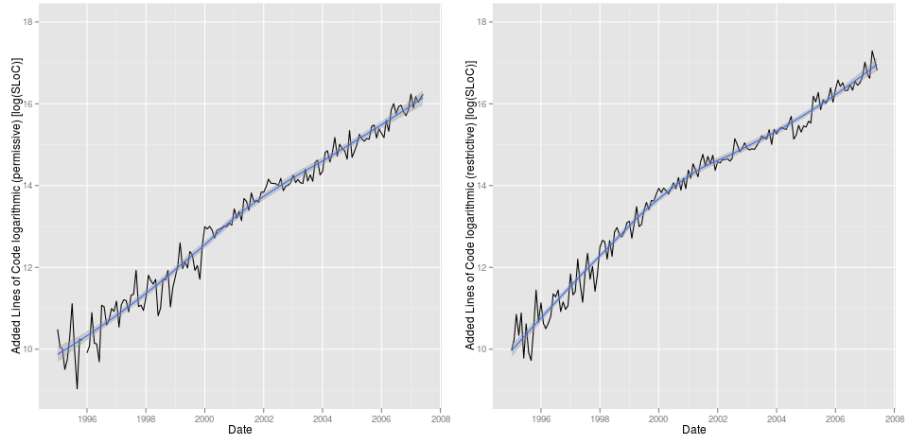

**Fig. 3.** Total SLoC added per month on log-scale

We found by visual inspection that for both sets there are two distinct periods of growth with a changing-point around 2000 to 2002. We estimated the points by conducting a segmented linear regression. The results are summarized in Table 2.

**Table 2.** Estimated changing-points

| License-type | Estimated changing-point | 95% Confidence | |
|---|---|---|---|
| | | 2.5% | 97.5% |
| Permissive | 2001-12 | 2000-06 | 2003-05 |
| Restrictive | 2000-02 | 1999-08 | 2000-08 |

The ordinary least-squares (OLS) estimator used for the linear regression is sensitive to autocorrelation in the data. We computed the Durbin-Watson-statistic[5] for both segmented linear models which returned significant autocorrelation at lag 1 for the permissive set and marginal autocorrelation at lag 1 for the restrictive set listed in Table 3.

**Table 3.** Autocorrelation and Durbin-Watson-Statistic for the segmented linear models up to lag 3

---

[5] The Durbin-Watson-statistic is approximately 2 for no autocorrelation. Values up to 0 or 4 indicate positive or negative autocorrelation [20].

| License-type | Lag | Autocorrelation | D-W-Statistic | p-value |
|---|---|---|---|---|
| Permissive | 1 | 0.197 | 1.560 | 0.002 |
|  | 2 | -0.086 | 2.117 | 0.600 |
|  | 3 | -0.076 | 2.093 | 0.590 |
| Restrictive | 1 | 0.137 | 1.725 | 0.062 |
|  | 2 | 0.038 | 1.913 | 0.492 |
|  | 3 | -0.020 | 1.944 | 0.670 |

To take the autocorrelation into account, for both models the two segments were re-fitted using the generalized least-squares (GLS) estimator which works as a maximum-likelihood-estimator even under the presence of correlation. The resulting fits are shown in Fig. 4.

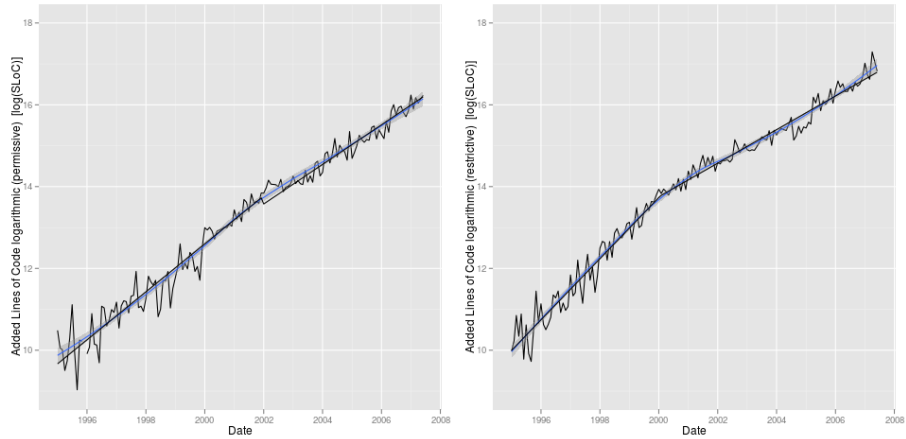

**Fig. 4.** Segmented linear models on log-scale of total added SLoC using GLS with blue Loess curve

The residuals are shown in Fig. 5 and the quantile-quantile (QQ)-Plots [20] in Fig. 6.

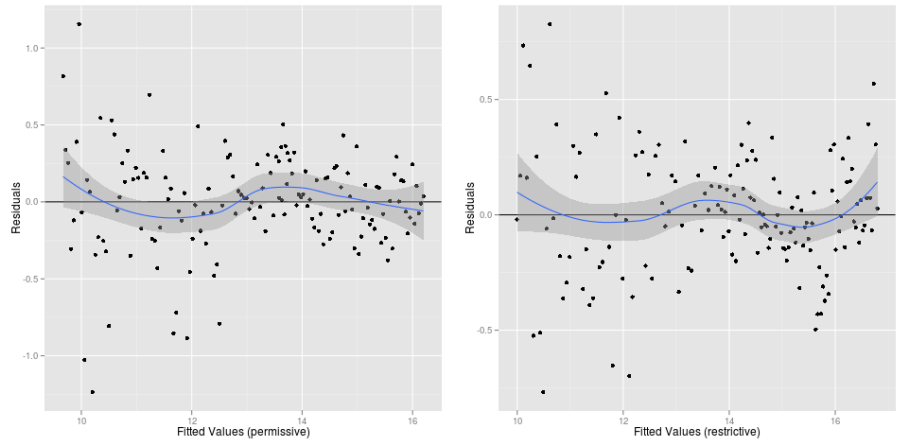

**Fig. 5.** Fitted values of segmented linear models using GLS on logarithmic data against residuals with Loess-curve in blue

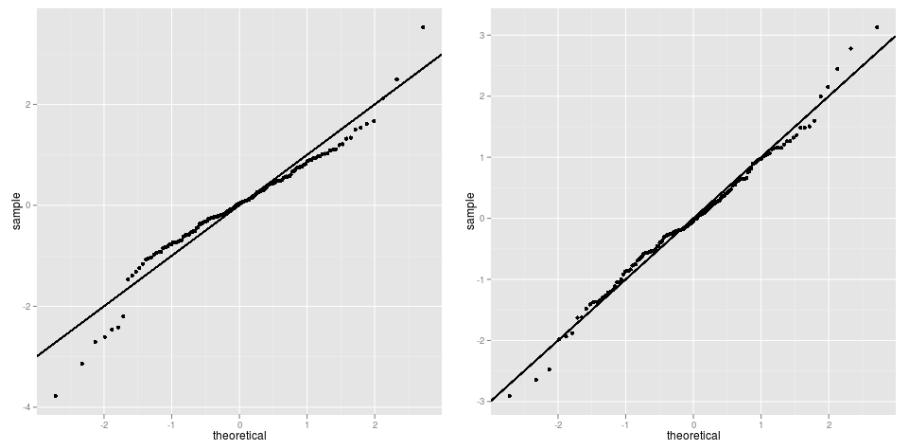

**Fig. 6.** QQ-plots of segmented linear models using GLS on logarithmic data.

Table 4 lists the slopes for both periods with 75% and 95% confidence intervals. During the first period, the restrictive set grows faster with a confidence of 95% while the trend reverses in the second period where the permissive set is growing faster with a confidence of 75%. Note that after the changing-point, both sets grow slower. But the restrictive set shows a stronger slowdown than the permissive one.

**Table 4.** Comparison of the slope of the segmented linear models using GLS on log-transformed response for the restrictive and permissive set including confidence intervals.

| Type | Period | Slope | 75% | | 95% | |
|---|---|---|---|---|---|---|
| | | | *12.5%* | *87.5%* | *2.5%* | *97.5%* |
| Perm. | 1 | 0.00160 | 0.00153 | 0.00167 | 0.00148 | 0.00172 |
| | 2 | 0.00133 | 0.00123 | 0.00143 | 0.00115 | 0.00150 |
| Restr. | 1 | 0.00205 | 0.00198 | 0.00210 | 0.00194 | 0.00215 |
| | 2 | 0.00112 | 0.00103 | 0.00122 | 0.00097 | 0.00128 |

An overview of the confidence levels regarding the differences-in-slopes is shown in Table 5:

**Table 5.** Confidence levels

| Period | Total Growth | Confidence |
|---|---|---|
| 1995-2001 | Restrictive > Permissive | > 95% |
| 2001-2007 | Restrictive < Permissive | 75% |

Beyond the changing-point, the different growth speeds can not be distinguished with 95% confidence, yet the results indicate that the initial trend was reversed and the permissive set has been growing faster since then.

Fig. 7 shows the models transformed to the original non-logarithmic scale. The restrictive model visually deviates from the Loess curve towards the end, an effect that is intensified by the high slope in that area. In the future the curves would intersect again.

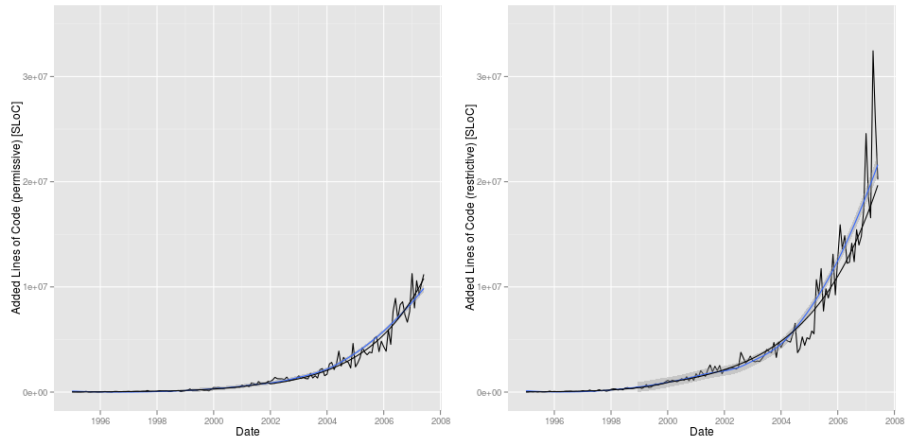

**Fig. 7.** Segmented linear models on normal scale

Table 6 lists the model fomulas.

**Table 6.** Models on nomal scale

| Type | Model[6] |
|---|---|
| Perm. | $y = 0.00694 \cdot e^{0.00160 \cdot x} \cdot e^{(-0.000277 \cdot (x-\psi)_t)} \cdot e^\varepsilon$ |
| Restr. | $y = 0.00017 \cdot e^{0.00205 \cdot x} \cdot e^{(-0.000922 \cdot (x-\psi)_t)} \cdot e^\varepsilon$ |

The back-transformed models include the error term, because the error roughly has a mean of zero for the linear models on the log-transformed response, which is no longer the case when the models get transformed back to normal scale. An estimate of the bias was conducted using the "smearing estimate of bias" method for residuals that are not normally distributed [21]. The bias needs to be taken into account when the models are used for prediction and is 1.049 (4.9%) for the permissive set and 1.035 (3.5%) for the restrictive one. We emphasize that this correction does, naturally, not eliminate the other complications associated with predictions from non-mechanistic models.

---

[6] $(x-\psi)_t$ defines a function where $\psi$ is the break-point and $(x-\psi)_t$ is 0 for $(x<\psi)$

## 5. Limitations

The quantitative analysis has shortcomings in regard to the database used:
- Sample size: The sample constituted of 1861 projects in the category 'permissive' and 3257 projects in the category 'restrictive'. The real number of active projects in both categories was much larger during the analyzed time-frame. Deshpande and Riehle [16] have estimated that the database holds roughly 30% of the active open source projects of the analyzed time-frame, a proportion we consider relevant to examine overall trends.
- Data incompleteness: The collection process began in 2005, a date where some open source projects had already discarded the earlier history, for example when moving to another RCS. However this does not affect the results regarding the differences in growth between the permissive and the restrictive set since the selection-bias does not differentiate between licenses.
- Project source: The snapshot of the Ohloh.net database had only connected to CVS, Subversion and Git source code repositories. Since almost all open source projects where maintained in one of these repositories during the analyzed time-frame this is only a minor limitation.
- Copy and paste: The database does not account for copy and paste. Copy and paste introduces a bias towards restrictively-licensed projects because a restrictive project can incorporate code from a permissive one but not vice-versa. To analyze the influence of this bias is a suggestion for further research.
- Aggregation: Also, we have only been looking at aggregate growth of open source projects, not at the growth of individual projects. We believe this to be justified, given the research question of this work. While some projects are large, the overall project size distribution has a long tail, making it impossible for any single project to have a substantial effect on the overall growth.
- Reproduceability: All data is publicly accessible and can be derived from the projects. The easiest way is to use the original data service, Ohloh.net, which has recently opened API access to its database for the general public.

## 6. Conclusions

This paper presents an empirical study of open source project growth using a large data set (about 30% of all active open source projects at its time). It repeats the prior finding that open source software code is growing at an exponential rate. It adds to that original finding a higher precision of the closed-form mathematical models of that growth. In addition, the paper looks at a project's open source license choice and provides a growth analysis binned by two dominant license types: permissive and restrictive (reciprocal) licenses. The paper provides analytically closed models for both license types and finds that both models are exponential as well. Surprisingly, both the permissively licensed and the restrictively licensed project data sets are best modeled by two separate exponential models with a changing point at around 2001 for both types of projects. Even more surprising, we find that restrictively licensed projects were growing faster than permissively licensed projects around until that changing point in 2001, and permissively licensed projects have been growing faster since then.

We attribute this finding to the growth of commercially sponsored open source communities, for example, the Linux, Apache, or the Eclipse Foundation [22]. Corbet found that most of the new code written for the Linux Kernel 2.6.20 was paid for by companies [23]. Similarly, in other yet unpublished work we have found an increasing and broad investment of company resources into community-owned open source. Such investments into a common good only make economic sense, if companies can reap benefits through complementary products that build on the common good. A restrictive license would restrict the creation of a competitively differentiated complementary product, so we believe that most companies will prefer a permissive license for the common good. The combined effect of increased commercial investment with the need for competitively differentiated products built on top of that shared investment has lead to an increase of permissively licensed projects and this obviously to such an extent, that number and size of permissively licensed projects have overtaken those of restrictively licensed projects. From this argument, we can only expect this trend to accelerate.

## References

1. Lerner, J., Tirole, J.: Some simple economics of open source. In: The journal of industrial economics, vol. L, no. 2 (2002)
2. Aksulu, A., Wade, M.: A comprehensive review and synthesis of open source research. October, vol. 11, no. 11, pp. 576-656 (2010)
3. Free Software Foundation, http://www.fsf.org
4. GNU General Public License v1, http://www.gnu.org/licenses/gpl-1.0.html
5. González, A. G.: Viral contracts or unenforceable documents? Contractual validity of copyleft licences. In: European Intellectual Property Review, pp. 1-20 (2004)
6. MIT License, http://www.opensource.org/licenses/MIT
7. BSD License, http://www.opensource.org/licenses/bsd-license.php
8. Sen R., Subramaniam C., Nelson M. L., "Determinants of the choice of open source software license," Journal of Management Information Systems, vol. 25, no. 3, pp. 207-240, Dec. 2008.
9. Lerner, J., Tirole J.: The scope of open source licensing. In: Source, vol. 21, no. 1 (2005)
10. Colazo, J., Fang, Y.: Impact of License Choice on Open Source Software Development Activity. In: Journal of the American Society for Information Science, vol. 60, no. 5, pp. 997-1011 (2009)

11. Aslett, M.: The trend towards permissive licensing, http://blogs.the451group.com/opensource/2011/06/06/the-trend-towards-permissive-licensing/ (2011)
12. Aslett. M.: FLOSSmole data confirms declining GPL usage, http://blogs.the451group.com/opensource/2011/06/13/flossmole-data-confirms-declining-gpl/ (2011)
13. Aslett. M.: On the continuing decline of the GPL, http://blogs.the451group.com/opensource/2011/12/15/on-the-continuing-decline-of-the-gpl/ (2011)
14. Ohloh, the open source network, http://www.ohloh.net/
15. FLOSSmole: Collaborative collection and analysis of free/libre/open source project data, http://flossmole.org/
16. Deshpande A., Alto P., Riehle D.: The Total Growth of Open Source, Source, no. 2006, p. 3 (2008)
17. Luckey, R.: The world's oldest source code repositories, http://meta.ohloh.net/2007/08/worlds_oldest_source_code_repositories/ (2007)
18. Koch S.: Evolution of open source software systems – A large-scale investigation, International Conference on Open Source Systems (2005)
19. Herraiz I., Gonzalez-Barahona J. M., Robles G., Rey U., Carlo J., "Towards a theoretical model for software growth," Fourth International Workshop on Mining Software Repositories (MSR'07) (2007)
20. Fahrmeir L., Regression - Modelle, Methoden und Anwendungen. Springer Verlag, (2009)
21. Newman M. C.: Regression analysis of log-transformed data: Statistical bias and its correction, Environmental Toxicology and Chemistry, vol. 12, no. 6, pp. 1129–1133 (1993)
22. Riehle D.: The Economic Case for Open Source Foundations. IEEE Computer vol. 43, no. 1 (January 2010). Page 86-90 (2010)
23. Corbet J.: Who wrote 2.6.20?, http://lwn.net/Articles/222773/